\documentclass[letterpaper,twocolumn,10pt]{article}
\usepackage{iftex}
\ifXeTeX  %
  \expandafter\def\csname ver@microtype.sty\endcsname{2099/01/01}
  \expandafter\def\csname opt@microtype.sty\endcsname{kerning,spacing}
  \expandafter\def\csname ver@breakurl.sty\endcsname{2099/01/01}  %
  \AtBeginDocument{}
\fi
\usepackage[hidelinks]{hyperref}  %
\usepackage{xurl}
\usepackage{usenix2019_v3}
\hypersetup{hidelinks}  %
\usepackage{amsmath,amssymb}
\usepackage{booktabs}
\usepackage{array}
\usepackage{graphicx}
\usepackage[export]{adjustbox}
\usepackage{xcolor}
\usepackage{xspace}
\usepackage{subcaption}
\usepackage{algorithm}
\usepackage[noend]{algpseudocode}
\usepackage{titlesec}
\titlespacing*{\section}{0pt}{8pt plus 2pt minus 1pt}{4pt plus 1pt}
\titlespacing*{\subsection}{0pt}{6pt plus 2pt minus 1pt}{3pt plus 1pt}
\titlespacing*{\paragraph}{0pt}{3pt plus 1pt}{0.6em}
\newif\ifshowsim\showsimfalse

\newcommand{\simtag}{\ifshowsim\textcolor{orange}{[SIM]}\fi\xspace}
\newcommand{\sys}{\textsc{Basin}\xspace}  %
\usepackage{tikz}
\usetikzlibrary{arrows.meta,positioning}
\definecolor{boxbg}{gray}{0.93}
\newcommand{\insightbox}[2]{\par\smallskip\noindent\fcolorbox{black!45}{boxbg}{\parbox{\dimexpr\columnwidth-2\fboxsep-2\fboxrule\relax}{\small\textbf{#1.} #2}}\par\smallskip}
\newcounter{observation}
\newcommand{\observation}[1]{\refstepcounter{observation}\insightbox{Observation~\theobservation}{#1}}

\begin{document}
\date{}
\title{\Large \bf When Cache Pressure Becomes a Trap: Metastability in Agent Serving}
\hypersetup{pdftitle={When Cache Pressure Becomes a Trap: Metastability in Agent Serving}, pdfauthor={Chengzhi Lu, Boxi Zhou, Yanying Lin, Dmitrii Ustiugov}}
\author{
\begin{tabular}{@{}c@{\hspace{8pt}}c@{\hspace{8pt}}c@{\hspace{8pt}}c@{}}
{\rm Chengzhi Lu} & {\rm Boxi Zhou} & {\rm Yanying Lin} & {\rm Dmitrii Ustiugov} \\[2pt]
\parbox[c][26pt][c]{110pt}{\centering\normalsize\normalfont Nanyang Technological University} &
\parbox[c][26pt][c]{110pt}{\centering\normalsize\normalfont Nanyang Technological University} &
\parbox[c][26pt][c]{110pt}{\centering\normalsize\normalfont Harvard University} &
\parbox[c][26pt][c]{110pt}{\centering\normalsize\normalfont Nanyang Technological University} \\
{\normalsize\normalfont Singapore} &
{\normalsize\normalfont Singapore} &
{\normalsize\normalfont Cambridge, MA, USA} &
{\normalsize\normalfont Singapore} \\[1pt]
{\footnotesize\normalfont\href{mailto:chengzhi.lu@ntu.edu.sg}{\texttt{chengzhi.lu@ntu.edu.sg}}} &
{\footnotesize\normalfont\href{mailto:bzhou011@e.ntu.edu.sg}{\texttt{bzhou011@e.ntu.edu.sg}}} &
{\footnotesize\normalfont\href{mailto:yanying_lin@seas.harvard.edu}{\texttt{yanying\_lin@seas.harvard.edu}}} &
{\footnotesize\normalfont\href{mailto:dmitrii.ustiugov@ntu.edu.sg}{\texttt{dmitrii.ustiugov@ntu.edu.sg}}}
\end{tabular}
}
\maketitle

\begin{abstract}
LLM agents repeatedly send a growing conversation to a model, making prefix reuse central to efficient serving. We show that a brief surge can disrupt this reuse and trigger \emph{metastable failure} in prefill--decode (PD) disaggregated engines: service remains degraded after arrivals return to a load the engine previously sustained. On SGLang, a ten-minute surge reduced goodput by about 88\% over the following hour. Removing the residual burst sessions restored service within minutes.

The sustaining mechanism is population--cache feedback. Slower calls prolong sessions and enlarge their aggregate context, or \emph{working set}; the resulting cache misses increase recomputation and further delay departures. A fluid model identifies the resource conditions for this feedback and explains how cache capacity, tool time and base load affect vulnerability. Across 36 simulated deployments, it flags all 17 cache-first cases. Controlled simulations hold the post-surge state fixed and show that changing subsequent arrivals alone reverses the recovery outcome. Once the cache collapses, low-reuse departure capacity sets the scale of the load reduction needed for recovery.

Guided by these dynamics, we build \sys, which gives each session a fixed priority, reserves their predicted peak contexts against an adaptive admission budget, extends the cache with idle decode memory, and sheds excess demand. On real H200 1P3D deployments under surges, including long-context SWE-bench Pro sessions, an arrival-time prototype achieves 1.20--1.70$\times$ the goodput of Continuum and 2.35--3.59$\times$ that of HiCache alone. In simulation, the predictor-driven policy delivers 1.20--2.05$\times$ the strongest baseline's goodput across six surge scenarios.

\end{abstract}

\section{Introduction}
\label{sec:intro}

\begin{figure}[t]
  \centering
  \includegraphics[width=\columnwidth]{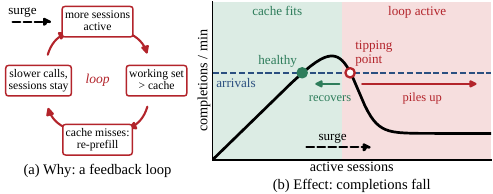}
  \caption{Population--cache feedback (schematic). (a) Cache misses extend session lifetimes, increasing the
  working set and further reducing reuse. (b) Departures rise with active sessions until cache pressure reduces
  throughput. A surge crossing the unstable intersection with base arrivals leaves departures below arrivals
  and active sessions accumulating.}
  \label{fig:concept}
\end{figure}

LLM agents build long conversations as they alternate between model calls and tool executions. Each call resends much of the preceding context: in the Terminal-Bench traces we study, the median prompt contains 25.6K tokens, 98\% of which appeared in the preceding call. Prefill--decode (PD) disaggregated engines~\cite{distserve,splitwise,mooncake} use prefix caching to reuse this earlier computation. A returning call then prefills only its new tokens, reducing prefill work by up to 50$\times$ when the prefix is available.

\paragraph{The problem.} This reliance on prefix reuse also creates a vulnerability: a brief surge can leave the engine persistently degraded even after arrivals return to a rate it previously sustained. On SGLang, a ten-minute burst reduces goodput by about 88\% over the following hour (\S\ref{sec:problem}). Before the burst, prefill and decode both operate well below saturation. Utilization, queue length and KV usage, commonly used for scaling and overload control~\cite{tokenscale,blitzscale,dynamollm,mooncake}, all show ample headroom and give little warning of the failure. The degradation persists after the trigger ends, a defining feature of \emph{metastable failure}~\cite{bronson2021metastable,huang2022metastable}.

\paragraph{Why it happens.} The degradation outlasts the burst because slower calls keep sessions active and competing for the cache. A surge enlarges the engine's \emph{working set}: the combined context of active sessions, including those executing tools. Cache contention evicts prefixes, forcing returning calls to recompute more tokens. The extra prefill work slows calls and keeps sessions active longer, since a session ends only after its final turn. New arrivals during this delay enlarge the working set, causing more eviction and recomputation (Fig.~\ref{fig:concept}). To test whether these accumulated sessions sustain the failure, we terminate the sessions introduced by the burst; service recovers within minutes. In a matching run that lets those sessions finish, active sessions keep accumulating after the last burst session departs, with arrivals at the original base rate.

Once this feedback develops, the engine can complete fewer sessions than arrive at the original base rate, so active sessions continue to accumulate after the surge ends. Recovery requires completions to exceed new arrivals, making the outcome depend on both the state left by the surge and the load that follows it. Controlled simulations confirm this dependence: runs with the same post-surge state recover under lighter arrivals and remain trapped under heavier arrivals. For a given deployment, the recovery boundary therefore depends on the base load.

\begin{figure}[t]
  \centering
  \includegraphics[width=\columnwidth]{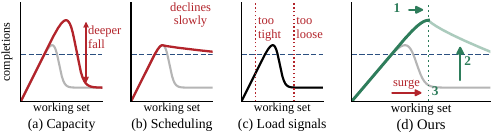}
  \caption{Existing remedies (a--c) and our design (d) on the curve of Fig.~\ref{fig:concept}b (gray: unchanged engine).
  (a) More capacity moves the cliff; vulnerability remains if the cache still limits the engine first. (b) Serving sessions already in progress first limits entry into the working set, but a surge can leave a long queue. (c) A fixed load or capacity threshold can admit too much or hold sessions out unnecessarily. (d) \sys combines fixed session priority (2), lent cache capacity (1), and admission and shedding for excess load (3).}
  \label{fig:remedies}
\end{figure}

\paragraph{Existing remedies.} To improve the balance between session arrivals and completions, existing systems change capacity, service order or admission. \textbf{Capacity expansion} adds cache tiers or borrows idle decode resources to raise prefill throughput~\cite{sglangHiCache,lmcache,mooncake,crossflow}. More cache supports more active sessions before the cliff, but feedback remains possible if cache pressure still limits service before decode saturates (Fig.~\ref{fig:remedies}a). \textbf{Scheduling} changes request or session priorities to preserve reuse and accelerate completion~\cite{sglang,continuum,infercept}. Prioritizing sessions already in progress can preserve reuse but leave new sessions waiting when completions barely exceed arrivals (Fig.~\ref{fig:remedies}b). \textbf{Admission control} bounds entry using KV usage, cache feedback or tier capacity~\cite{concur,thunderagent}. KV usage can remain low while tool-idle sessions accumulate prefix demand, whereas a fixed cache-capacity budget can delay admission even while reuse remains high (Fig.~\ref{fig:remedies}c).

\paragraph{Our approach.} To explain how capacity, scheduling and admission change population--cache feedback, we model the active session population as a fluid, with departures determined by closed-loop demand, cache reuse and decode capacity (\S\ref{sec:model}). The model identifies deployments where cache pressure limits service before decode saturates, flagging all 17 cache-first cases in a 36-deployment comparison. It explains how cache capacity, tool time and base load change vulnerability, and why preserving reuse can still leave a long backlog. Recovery also depends on cache history: in simulation, hit rates remain low after collapse at working sets that sustain high reuse in healthy runs. This hysteresis limits throughput during recovery.

We use this analysis to design \sys, which tracks session contexts across calls and tool executions (Fig.~\ref{fig:remedies}d). It assigns each session a fixed priority from its arrival time and predicted peak context, using the same order for admission and both engine queues. Idle decode memory preserves prompts between calls, reducing recomputation. Running decode requests reclaim the lent memory on demand (\S\ref{sec:design}). An adaptive working-set budget regulates admission, and shedding removes persistent excess demand.

On real H200 1P3D deployments, an arrival-time prototype delivers up to 3.59$\times$ HiCache's goodput and 1.70$\times$ that of the strongest baseline, Continuum. In simulation with the working-set predictor, \sys achieves the highest goodput in all six end-to-end scenarios, 1.20--2.05$\times$ the strongest baseline's. In the simulated $3$--$6\times$ surge-size sweep on 1P3D, \sys avoids trapping and completes at least 95\% of sessions while shedding at most 2.2\% (\S\ref{sec:evaluation}).

This paper makes three contributions:
\begin{itemize}
\item We identify population--cache feedback: slower calls keep sessions active, and continuing arrivals sustain cache thrashing after a finite surge. Real-system interventions establish its causal role in metastable failure in PD-disaggregated agent serving (\S\ref{sec:problem}).
\item We develop a fluid model that identifies resource conditions for cache-driven vulnerability and explains the effects of capacity, load and entry control. Across 36 simulated deployments, it flags every cache-first case. Controlled interventions reveal how continuing arrivals and cache hysteresis govern recovery (\S\ref{sec:characterization}, \S\ref{sec:model}).
\item We design \sys to preserve reuse through a fixed session priority, admission by predicted peak context, and decode-memory lending. A SGLang prototype establishes the benefits of session ordering and lending on hardware; simulation evaluates the full predictor-driven policy across deployments and loads (\S\ref{sec:design}, \S\ref{sec:evaluation}).
\end{itemize}

\section{Background}
\label{sec:background}

\paragraph{Prefill--decode disaggregation.} Each model call begins with prefill, which processes the prompt in parallel, followed by decode, which generates one token per request at each step. These phases have different compute and memory demands, so PD engines place them on separate instances and transfer the request's key--value (KV) state from prefill to decode~\cite{distserve,splitwise,mooncake}. In SGLang, a request first reserves KV space on decode, then runs prefill, transfers its state and begins decoding. Decode retains this KV throughout generation, often requiring several decode instances per prefill instance. Reported deployments range from 2.25 and 3 decode nodes per prefill node for DeepSeek and Kimi K2 on SGLang~\cite{sglangEP,sglangK2} to a 4-node prefill unit paired with a 40-node decode unit in DeepSeek's deployment~\cite{deepseekv3}.

\paragraph{Prefix caching.} The KV state computed during prefill can also be reused by later calls that share the same prefix. SGLang retains prompt KV in a radix tree on prefill, allowing a request that extends a cached prefix to prefill only the uncached suffix~\cite{sglang}. When memory fills, the engine evicts least-recently-used prefixes. Hierarchical caches such as HiCache extend capacity with finite host-memory or remote-storage tiers~\cite{sglangHiCache,mooncake}. Hybrid models also maintain fixed-size recurrent state in Mamba, lightning or gated-delta layers~\cite{jamba,minimax01,qwen36}. Reusing a prefix requires retaining this state alongside its KV, so recurrent-state slots impose another capacity constraint.

\paragraph{Agentic workloads.} Agents revisit these prefixes as they alternate between model calls and tool executions. In the ReAct pattern, each action's result is appended to the context before the next call~\cite{react}, often for tens of turns in coding tasks~\cite{sweagent,swebench,terminalbench}. Serving systems expose these sequences as programs that pause during tool execution~\cite{infercept,autellix,continuum}. Their contexts remain useful for subsequent calls even while the tools are running~\cite{aries}. The resulting workload is \emph{partly-open}~\cite{schroeder2006open}: new sessions arrive independently of engine progress, but each session issues its next call only after the preceding call and tool execution finish. Slower service therefore delays subsequent calls and keeps sessions active longer, while new sessions continue to arrive.

\paragraph{Metastable failures.} Slow service keeps more sessions active, and their increased cache demand can, in turn, sustain the slowdown through additional cache misses. This follows the pattern of \emph{metastable failure}: a trigger drives a vulnerable system into a degraded state, and a sustaining effect preserves that state after the trigger ends~\cite{bronson2021metastable}. Retries, garbage collection and cold look-aside caches are known sustaining effects~\cite{huang2022metastable}.

\section{Metastable Failures in Agent Serving}
\label{sec:problem}

\subsection{Workload and Setup}

\begin{figure}[t]
  \centering
  \includegraphics[width=\columnwidth]{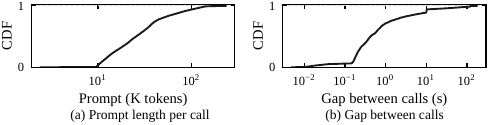}
  \caption{Workload: 216 Terminal-Bench trajectories, 7{,}108 calls.}
  \label{fig:workload}
\end{figure}

Our real-system experiments replay 216 Terminal-Bench trajectories containing 7{,}108 calls. Prompts grow incrementally: the median contains 25.6K tokens and the 90th percentile 77K, with 98.2\% of prompt tokens repeated from the preceding call (Fig.~\ref{fig:workload}a). Outputs are short, at a median of 160 tokens, and the median gap from a response to the next call is 0.42\,s (Fig.~\ref{fig:workload}b). These sessions reuse their contexts for minutes, making prefill cost sensitive to whether their prefixes remain cached.

We serve this workload with Qwen3.6-35B-A3B (FP8), a hybrid linear-attention MoE model, using SGLang in PD disaggregation on one prefill and one decode instance, each on an H100 NVL GPU and connected by NIXL. The prefill instance holds a 1.09M-token prefix cache and the decode instance 1.66M tokens of KV. Sessions arrive as a Poisson process and replay their recorded calls and tool gaps. Each run starts from steady operation with 42 active sessions joined mid-trajectory.

\subsection{A Transient Surge, a Persistent Failure}
\label{sec:problem-surge}

\begin{figure}[t]
  \centering
  \includegraphics[width=\columnwidth]{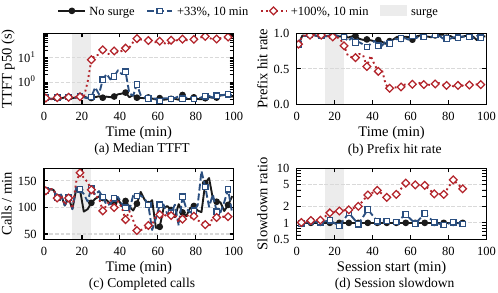}
  \caption{Same arrivals at a base load of 3.3 sessions/min, with and without a surge in minutes
  15--25 (shaded). (d) compares sessions by start time against the no-surge run.}
  \label{fig:surge}
\end{figure}

The engine can remain degraded long after a surge ends, even at a base rate it previously sustained. Fig.~\ref{fig:surge} compares the same base arrival stream at 3.3 sessions/min, with and without ten-minute surges of $+33\%$ and $+100\%$. Without a surge, the engine uses a twelfth of its saturated prefill throughput and maintains a median time to first token (TTFT) of about 0.2\,s. It recovers from the $+33\%$ surge within twenty minutes. The $+100\%$ surge, however, leaves the engine deteriorating after arrivals return to 3.3 sessions/min: median TTFT reaches about 60\,s and prefix hit rate falls from 0.95 to 0.27. From minute 25 onward, it completes 0.80$\times$ as many calls as the surge-free run. Under the goodput definition of \S\ref{sec:eval-method}, sessions arriving in the following hour achieve 17 sessions/h against 138 without the surge, and goodput over the whole run falls from 123 to 43 sessions/h. Sessions arriving after the surge take 1.8--6.1$\times$ longer (Fig.~\ref{fig:surge}d), and the degradation persists for the remaining hour.

\subsection{Residual Sessions Sustain the Failure}
\label{sec:problem-population}

\begin{figure}[t]
  \centering
  \includegraphics[width=\columnwidth]{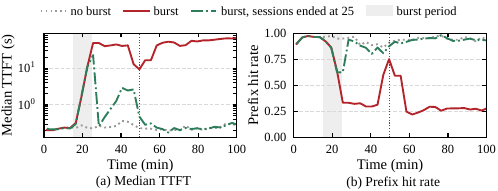}
  \caption{The same arrivals at a base load of 3.3 sessions/min: without a burst, with a burst of short sessions
  at 4$\times$ the base rate in minutes 15--25 (shaded), and with the same burst whose sessions all end at
  minute 25. Dotted line: last call of the burst sessions that are left in place.}
  \label{fig:persist}
\end{figure}

To test whether sessions left active after a burst sustain the degradation, we inject short sessions at four times the base rate during minutes 15--25. These sessions complete within five minutes on a healthy engine. We compare two runs: one lets the burst sessions finish on their own, and the other terminates them all at minute 25. Both runs retain the same base arrivals.

\begin{figure}[t]
  \centering
  \includegraphics[width=\columnwidth]{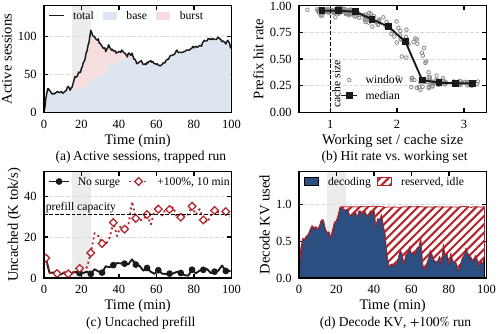}
  \caption{The sustaining loop in real runs; shaded: the surge. (a) Active sessions after a burst of short sessions at
  4$\times$ the base rate, split into base and burst sessions. (b) Each 2-minute window of all runs
  and their binned median; dashed: cache size. (c) Prefill tokens that miss the cache. (d) Decode KV of decoding
  requests vs.\ reserved for requests awaiting prefill.}
  \label{fig:mechanism}
\end{figure}

By the end of the burst at minute 25, both runs have about 110 active sessions, including 65 burst sessions, and median TTFT is about 9\,s. Terminating the burst sessions restores service within two minutes. When left to finish on their own, the burst sessions remain active for up to 25 more minutes, while the number of active base sessions grows to 1.7$\times$ that in the no-burst run by minute 50. The hit rate recovers only briefly when the last burst sessions finish. Active base sessions continue to accumulate, reaching 2.4$\times$ the no-burst count by minute 90 (Figs.~\ref{fig:persist} and~\ref{fig:mechanism}a).

Recovery after removing the burst sessions shows that the accumulated sessions sustain the failure. At the same base arrival rate, the engine either remains healthy or degrades persistently depending on its earlier exposure to the burst. This dependence on past load is a defining feature of metastable failure (\S\ref{sec:background}).

\subsection{Population--Cache Feedback}
\label{sec:problem-loop}

\paragraph{Session lifetime.} Slower calls keep sessions active longer because each session departs only after its final turn. Meanwhile, new arrivals add more active sessions. In steady state, Little's law relates mean population to arrival rate $\lambda$ and mean lifetime $T$ as $\lambda T$, where $T$ includes serving and tool time.

\paragraph{Cache pressure.} As sessions accumulate, so does their \emph{working set}: the aggregate context of all active sessions, including those executing tools. The growing working set causes LRU to evict prefixes before their sessions return. Across the two-minute windows of our real runs, hit rate declines from 0.95 to 0.80 before dropping sharply to about 0.3 when the working set reaches roughly twice the cache capacity (Fig.~\ref{fig:mechanism}b). We call this transition the \emph{cache cliff}. The resulting recomputation raises uncached prefill work in the $+100\%$ run from 2.5K to 28--35K tokens/s, near the saturated throughput of 31K tokens/s (Fig.~\ref{fig:mechanism}c).

Prefill delays also tie up decode memory because SGLang reserves decode KV before prefill begins. After the $+100\%$ surge, requests awaiting prefill hold two thirds of decode KV (Fig.~\ref{fig:mechanism}d). Decode memory is 97\% full, yet only half as many requests decode as in the surge-free run. New requests wait 25\,s for a reservation, further extending session lifetimes.

Together, these effects form a feedback loop: cache misses prolong sessions, and the additional active sessions sustain cache misses. The $+33\%$ surge leaves the working set below the cliff and the engine recovers. The $+100\%$ surge crosses it, reducing departures below base arrivals. In the short-session experiment, removing the burst sessions reduces the working set and breaks this loop. The cliff raises service cost sharply; recovery then depends on whether the engine can complete sessions faster than new ones arrive. Section~\ref{sec:model} models this balance and tests the role of continuing arrivals.

\observation{A transient surge can sustain prefix-cache thrashing by extending session lifetimes and enlarging the working set. Removing the residual burst sessions breaks this feedback; recovery depends on both the working set and the base arrival rate.}

\section{Characterizing Vulnerability}
\label{sec:characterization}

The feedback identified in \S\ref{sec:problem} creates a vulnerable region below steady-state overload. We map this region and show how deployment resources change it using a calibrated discrete-event simulator. Plots in this section and \S\ref{sec:model} report simulation results unless explicitly identified as real measurements.

\paragraph{Methodology.} The simulator replays the real driver's trajectories, arrival process and warm start. Each call is released only after its predecessor completes and the recorded tool gap elapses. Slower service therefore keeps sessions active longer while new sessions continue to arrive, preserving the session-level feedback. The simulator models prefill queues and cache tiers, decode KV reservation, and decode costs as functions of batch size and running context. H100 and H200 use separate engine calibrations (\S\ref{sec:eval-sim}).

We summarize persistent latency in these sweeps using $L_{\mathrm{late}}$, the maximum reported window-median TTFT in the final 30 minutes. A value above 10\,s identifies persistent high latency. The controlled post-surge intervention in \S\ref{sec:model-boundary} uses the full trapped-state definition of \S\ref{sec:eval-method}.

\subsection{Vulnerability Below Saturation}

\begin{figure}[t]
  \centering
  \includegraphics[width=\columnwidth]{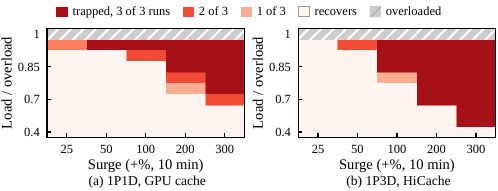}
  \caption{\simtag How often a ten-minute surge traps the engine, by base load (relative to
  steady-state overload) and surge size.}
  \label{fig:map2d}
\end{figure}

Finite surges trigger persistent cache thrashing well below steady-state overload. Following prior metastability studies~\cite{huang2022metastable}, we vary base load and the size of a ten-minute surge, from $+25\%$ to $+300\%$. On 1P3D with HiCache, a $+300\%$ surge traps the engine at half the overload rate measured without a surge (Fig.~\ref{fig:map2d}). As base load approaches that rate, the surge needed to trigger failure decreases to $+50\%$ on 1P3D and $+25\%$ on GPU-only 1P1D. Trigger sizes in this range also occur in production arrivals: after thinning two Alibaba Bailian traces~\cite{kvcachewild} to a single deployment's rate, their busiest ten-minute windows exceed the hourly mean by 16--49\%.

\subsection{More Cache Moves the Boundary}

\begin{figure}[t]
  \centering
  \includegraphics[width=\columnwidth]{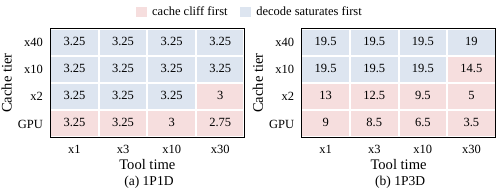}
  \caption{\simtag Highest load (sessions/min) that recovers from a $+100\%$ surge, by cache tier
  (GPU only, or a host tier of 2--40$\times$ the GPU cache) and tool time. Color: the first resource
  to limit service.}
  \label{fig:boundaries}
\end{figure}

Adding cache raises the recoverable load until decode becomes the first limit. Fig.~\ref{fig:boundaries} varies HiCache host capacity, the ratio of prefill to decode instances (P:D), and tool time. For each configuration, we report the highest base load that recovers from a $+100\%$ surge and the limiting resource above that load. On 1P3D at the default tool time, expanding from GPU-only caching to a host tier 40 times GPU capacity raises this recoverable load from 9 to 19.5 sessions/min.

More decode instances and longer tool execution both shift the bottleneck toward the prefix cache. Additional decode instances expand generation capacity relative to the prefill cache, while tool execution prolongs demand for cached contexts between decode requests. On 1P1D with a host tier, decode is the first observed bottleneck except at the longest tool times. On 1P3D, cache reuse deteriorates before decode saturates with either GPU-only caching or a host tier twice the GPU cache at every evaluated tool time. With tool time scaled by 30, the cache remains the first limit even with a host tier ten times the GPU cache. A trapped deployment with a 125\,GB host tier reaches a median TTFT of 335\,s, 3.3$\times$ the GPU-only H100 value.

Larger GPUs raise the absolute triggering load while leaving substantial vulnerability below overload. The H200 profiles provide roughly twice the cache and decode memory and 1.6--1.75$\times$ the nominal prefill throughput of H100. On 1P3D, the first load trapped by a $+300\%$ surge rises from 3.5 sessions/min on H100 to 5.5 on H200. Across the GPU and host-tier sweep, this trapping load remains between 0.46 and 0.70 of the steady-state overload rate. Proportional expansion of prefix-cache and decode memory preserves their ratio. The model in \S\ref{sec:model} relates this ratio to the fraction of active sessions in service to identify which resource limits the engine first.

\observation{Finite surges trigger cache-driven failure well below steady-state overload. More cache raises the recoverable load; additional decode capacity and longer tool execution shift the first limit toward the cache.}

\section{A Fluid Model of Session Dynamics}
\label{sec:model}

The model tracks how arrivals and completions change the active session population. It connects this balance to prefix reuse and prefill--decode capacity, explaining the vulnerable regions in \S\ref{sec:characterization} and how entry control and added resources change them.

\subsection{Population and Resource Limits}
\label{sec:model-fluid}

\paragraph{Population dynamics.} Let $N(t)$ count active sessions, including those executing tools; we call this the \emph{session population}. Initially, assume every arriving session becomes active. Sessions arrive at rate $\lambda(t)$ and issue $K$ calls on average, separated by tool gaps of mean $g$. We approximate the session departure rate by call throughput divided by calls per session:
\begin{equation}
  \frac{dN}{dt} \;=\; \lambda(t) - \frac{X(N)}{K},
  \label{eq:fluid}
\end{equation}
where $X(N)$ is call throughput at population $N$. Throughput is constrained by closed-loop demand, prefill work and decode capacity; we combine their bounds with a smooth minimum. At small $N$, the closed loop limits throughput to $N/(g+L_0)$, where $L_0$ is unloaded call latency.

As cache misses increase prefill work, the prefill throughput bound $\mu_p/u(N)$ falls. Here $\mu_p$ is saturated prefill throughput and $u(N)=\max\bigl(n_{\text{new}},(1-h(W/C))P\bigr)$ is uncached work per call. The mean prompt length is $P$, and $n_{\text{new}}$ is the new suffix length. The hit rate $h(W/C)$ depends on working set $W=N\bar c$ relative to prefix-cache capacity $C$, where $\bar c$ is mean context length across active sessions. The model uses the real-system hit-rate curve in Fig.~\ref{fig:mechanism}b.

Decode capacity bounds throughput by the number of requests whose KV fits in memory divided by their decode time. Comparing this bound with the cache-dependent prefill bound determines whether falling reuse can reduce throughput before decode limits service.

\begin{figure}[t]
  \centering
  \includegraphics[width=\columnwidth]{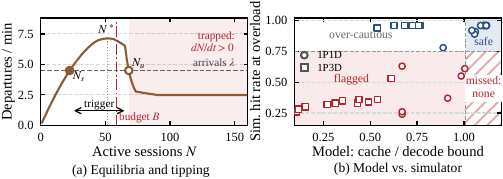}
  \caption{\simtag (a) Departure rate in the fluid model for 1P3D with HiCache: arrivals meet it at a stable equilibrium $N_s$ and an unstable one $N_u$, beyond which the population grows. (b) Resource-based vulnerability classification across simulated deployments. The model flags those left of the dashed line; all 17 cache-first deployments fall in this region.}
  \label{fig:hump}
\end{figure}

\paragraph{Equilibria and tipping.} When cache misses make prefill the first limit, the departure curve $D(N)=X(N)/K$ rises with the session population, peaks at $N^\ast$, and then falls. After reuse collapses, departures approach a low-reuse rate $D_{\text{low}}$. A base arrival rate between $D_{\text{low}}$ and the peak $D(N^\ast)$ intersects the curve twice: at a stable equilibrium $N_s$ and an unstable equilibrium $N_u$ (Fig.~\ref{fig:hump}a). Populations below $N_u$ return toward $N_s$; beyond $N_u$, departures remain below arrivals and the population grows. The same base load thus supports healthy service before a surge and sustained population growth afterward. Raising base load moves $N_s$ and $N_u$ closer together, so a smaller population increase triggers failure. Arrivals below $D_{\text{low}}$ let the population drain while reuse remains poor.

\paragraph{Vulnerability condition.} The cache cliff and decode-memory saturation impose competing limits on the active population. The measured cliff is near $N_c\approx1.9\,C/\bar c$. Let $D_{\text{KV}}$ be decode KV capacity and $f$ the fraction of active sessions in service. Decode memory then fills near $N_d\approx D_{\text{KV}}/(f\bar c)$. Comparing these population estimates gives the condition for the cache cliff to precede decode-memory saturation:
\begin{equation}
  C \;\lesssim\; \frac{D_{\text{KV}}}{2f}.
  \label{eq:condition}
\end{equation}
More host capacity raises $N_c$, moving the cache cliff to a larger population. More decode memory raises $N_d$, and longer tool gaps raise $N_d$ by reducing $f$: more sessions are executing tools at a given decode occupancy. These shifts explain why added decode capacity and longer tool execution favor cache-first failure. Scaling $C$ and $D_{\text{KV}}$ together preserves the criterion at fixed $f$; GPU service rates can also change $f$.

\paragraph{Deployment classification.} To account for service rates as well as memory, we compare the model's peak throughput under the closed-loop and prefill bounds with its decode throughput bound. Fig.~\ref{fig:hump}b plots their ratio for 36 simulated deployments spanning P:D ratios, cache tiers and tool times. With engine costs matched to the simulator and a 1\% tolerance at the crossover, the model flags all 17 cache-first deployments. It also flags eight of the 19 decode-first deployments, giving 28 correct classifications (78\%).

\begin{figure}[t]
  \centering
  \includegraphics[width=\columnwidth]{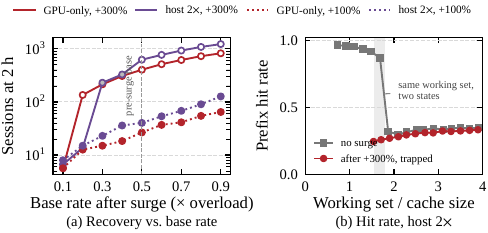}
  \caption{\simtag Post-surge arrival intervention on 1P3D with HiCache alone (three seeds). (a) Mean active sessions at the end of the two-hour run, by post-surge arrival rate; the dashed line marks the pre-surge rate. Filled markers: no seed trapped; hollow: all three trapped; gray: some trapped, using the state definition of \S\ref{sec:eval-method}. (b) Host-tier hit rate at comparable working sets, with and without a prior collapse. Trapped-run samples cover minutes 40--120; shading marks the overlap.}
  \label{fig:recbound}
\end{figure}

\subsection{Continuing Arrivals Sustain Failure}
\label{sec:model-boundary}

For a fixed post-surge state, continuing arrivals determine whether active sessions drain or keep growing. We isolate this effect by running HiCache alone on H200 1P3D with GPU-only caching or a host tier twice the GPU cache. Each run starts at half the deployment's reference overload rate $\lambda_{\mathrm{ref}}$ and receives a ten-minute surge in minutes 15--25. We then set arrivals to $0.1$--$0.9\,\lambda_{\mathrm{ref}}$. Within each seed and surge setting, the runs have identical histories and engine states through minute 25. We apply the trapped-state definition of \S\ref{sec:eval-method} to the final two 20-minute windows, comparing against no-surge controls with the same arrival schedule. The reference rates are 12.0 sessions/min with GPU-only caching and 16.5 with the host tier.

Changing only subsequent arrivals reverses the outcome (Fig.~\ref{fig:recbound}a). After the $+300\%$ surge on GPU-only 1P3D, every seed recovers at $0.1\,\lambda_{\mathrm{ref}}$ and remains trapped at $0.2\,\lambda_{\mathrm{ref}}$. With the host tier, every seed recovers at $0.2\,\lambda_{\mathrm{ref}}$ and remains trapped at $0.5\,\lambda_{\mathrm{ref}}$; two of three seeds are trapped at each of $0.3\,\lambda_{\mathrm{ref}}$ and $0.4\,\lambda_{\mathrm{ref}}$. After the smaller $+100\%$ surge, these same deployments remain healthy throughout the tested rates. Larger surges therefore leave states that require a greater reduction in arrivals to recover. Recovery depends jointly on the state left by the surge and the arrival rate that follows.

\paragraph{Cache history and recovery.} After cache collapse, reuse remains low even at working sets that support high hit rates in healthy runs (Fig.~\ref{fig:recbound}b). On host-tier 1P3D, the simulated hit rate remains at 0.26--0.37 after collapse at working sets 1.6--1.7 times cache capacity; no-surge runs at comparable working sets retain 0.81--0.93. This is cache hysteresis: similar demand for cached context produces different hit rates because the cached contents differ. Prefixes evicted during collapse must be recomputed before they become reusable again.

The low-reuse state limits the departure capacity available for recovery. In trapped runs, measured departures are 0.20--0.31 of $\lambda_{\mathrm{ref}}$ on GPU-only 1P3D and 0.20--0.24 with the host tier. At the pre-surge arrival rate of $0.5\,\lambda_{\mathrm{ref}}$, arrivals exceed these departures and keep sessions accumulating. Recovery requires a departure surplus while reuse remains low. Throughput measured in this collapsed state sets the scale of the arrival reduction needed to drain active sessions.

\subsection{Implications for Mitigation}
\label{sec:model-remedies}

In Eq.~\ref{eq:fluid}, every base rate between $D_{\text{low}}$ and the departure peak is vulnerable to a finite surge. Mitigations act on this balance by regulating entry into the active population and raising the departure curve.

\emph{Serving admitted sessions first controls entry.} Sessions waiting for their first call have yet to create private cached context. Let $Q(t)$ count these sessions and $I(t)$ be their admission rate into the active population. Then
\begin{equation}
  \frac{dN}{dt}=I(t)-D(N), \qquad
  \frac{dQ}{dt}=\lambda(t)-I(t).
  \label{eq:entry}
\end{equation}
Under resident-first scheduling, new sessions enter using the service capacity left after resident calls. As cache misses increase resident prefill work, less capacity remains for entry, opposing further population growth. This feedback explains how program ordering with KV retention across tool calls~\cite{continuum,infercept} preserves reuse while new sessions accumulate in an external queue.

When ordering maintains the population near the departure peak, $N\approx N^\ast$, the queue evolves as
\begin{equation}
  \frac{dQ}{dt} \;\simeq\; \lambda(t)-D(N^\ast).
  \label{eq:queue}
\end{equation}
At constant base load $\lambda<D(N^\ast)$, a backlog of $S$ sessions takes approximately $S/(D(N^\ast)-\lambda)$ to drain. The drain time therefore diverges as base arrivals approach the departure peak from below, even with healthy cache reuse.

\emph{Capacity raises the departure peak.} More cache supports a larger resident population before recomputation limits service. A host tier, a larger GPU or idle decode memory can supply this capacity. Borrowing decode compute for prefill~\cite{crossflow} raises the prefill bound directly. A falling departure branch leaves base rates between $D_{\text{low}}$ and the new peak vulnerable to finite surges after capacity expands. Once decode limits service, decode capacity caps the departure peak. Under resident-first order, the larger departure surplus also drains the external queue faster.

\emph{Admission explicitly bounds entry.} A working-set budget scaled to the cache cliff regulates admission according to the working-set principle~\cite{denning1968working}. Parking first calls when the budget is full regulates the entry rate $I(t)$ into the active population.

Admission divides unfinished sessions between $N$ and $Q$. Adding their balances in Eq.~\ref{eq:entry} gives $d(N+Q)/dt=\lambda-D(N)$. For a fixed departure curve and sustained $\lambda>D(N^\ast)$, this total grows at least at rate $\lambda-D(N^\ast)$. Bounding the backlog therefore requires added departure capacity or long-run shedding of at least this excess rate.

\edef\basinSavedTopNumber{\number\value{topnumber}}
\setcounter{topnumber}{1}

\section{Basin Design}
\label{sec:design}

\begin{table}[t]
\footnotesize
\centering
\setlength{\tabcolsep}{4pt}
\caption{Design principles and their realization in \sys. $W=N\bar c$: working set; $\hat W_i$: predicted peak context of session $i$; $W_c$: cliff; $C_{\text{eff}}$: cache capacity; $B_0=\min(mW_c,C_{\text{eff}})$: reference budget; $B$: adapted budget.}
\label{tab:control}
\begin{tabular}{@{}>{\raggedright\arraybackslash}p{0.13\columnwidth}>{\raggedright\arraybackslash}p{0.19\columnwidth}>{\raggedright\arraybackslash}p{0.24\columnwidth}>{\raggedright\arraybackslash}p{0.34\columnwidth}@{}}
\toprule
Control & Model term & Mechanism & Engaged when \\
\midrule
Priority & entry balance: Eq.~\ref{eq:entry} & fixed session priority & persistent prefill backlog, wait for decode memory, or $W \ge \alpha_O B_0$ \\
Capacity & $C$, hence $W_c$ and $D(N^\ast)$ & lend decode memory as a cache tier & $W \ge \alpha_L B_0$ of the prefill tiers or admission active, decode has spare memory \\
Entry & $N$ & Denning admission: $\sum\hat W_i\le B$ & $W \ge B_0$ \\
 & $Q$ when $\lambda > D(N^\ast)$ & shedding & long-run arrivals exceed the peak \\
\bottomrule
\end{tabular}
\end{table}

\sys follows three principles from the model: preserve reuse, raise departure capacity, and regulate entry. It applies Denning's working-set principle~\cite{denning1968working} at session granularity, reserving for each admitted session the largest context it is predicted to reach. Each session receives a fixed priority from its arrival time and predicted demand. This priority orders admission and both engine queues, favoring smaller sessions while limiting how long later arrivals can overtake earlier ones. Decode-memory lending extends the cache and raises departures on cache-limited deployments, while shedding removes persistent excess demand. Table~\ref{tab:control} maps these principles to the model terms they change and their activation conditions.

A session gateway runs the controller, and extensions to the prefill and decode engines apply its decisions. The gateway combines a deployment profile with measurements of outstanding prefill work $Q_P$, requests waiting for decode memory $Q_D$, prefix hit rate $h$, working set and long-run arrival rate $\bar\lambda$. It evaluates these signals every control interval $\Delta$, and more often while backlog or an active control persists. Releasing a control requires its condition to remain clear for a cool-down.

\subsection{Working-Set Estimation}
\label{sec:design-model}

A session's latest context remains reusable while its tools run, so \sys counts it throughout the session's lifetime. To decide when controls engage, it estimates aggregate demand as $W=N\bar c$, where $N$ counts active admitted sessions and $\bar c$ is a smoothed mean context length. New sessions immediately contribute $\bar c$ to this estimate, allowing the controller to anticipate their demand before their contexts fully develop.

Admission reserves demand per session. Since context grows across turns, the first prompt alone understates a session's eventual footprint. At the first call, \sys predicts $\hat W_i$, the session's peak context length, including input and output tokens. The predictor uses the task text and first-prompt length visible at that call (\S\ref{sec:design-profile}). The prediction stays fixed until completion, and the sum of these reservations over residents governs further admission. This sum accounts for future growth; $W=N\bar c$ supplies the live demand signal that activates controls.

The deployment profile gives a cliff $W_c=\kappa x_k C_{\text{eff}}$, where $x_k\approx1.9$ is the reference hit-curve knee and $\kappa$ scales this curve to the deployment. Let $C_G$ and $C_H$ denote aggregate prefill GPU and inclusive host capacity, and $L$ the estimated lent capacity. We use $C=\max(C_G,C_H)$ and $C_{\text{eff}}=\max(C,C_G+L)$ to account for overlap between tiers. The prototype estimates $L$ by smoothing the capacity of retained units.

The reference budget is $B_0=\min(mW_c,C_{\text{eff}})$, combining a margin $m<1$ below the profiled cliff with the physical cache capacity. The capacity term anchors reservations to available memory; the cliff term accounts for locality lost through routing or fragmentation. Counting each session's peak separately reserves space for its later growth, while shared prefixes and tool-idle intervals can make actual occupancy smaller. Hit-rate feedback adapts the admission budget $B$ around $B_0$ to exploit this reuse. Thus $B_0$ provides a reference for control, and the current $B$ sets the reservation limit when admission is active.

The profile also supplies the peak departure rate $\lambda_c$ used for shedding.

\subsection{Memory Lending for Prefix Reuse}
\label{sec:design-capacity}

Lending expands the cache with idle decode memory while prefill computation remains on prefill instances. On cache-limited deployments, retaining more prefixes reduces recomputation and raises departure capacity. \sys enables lending once $W\ge\alpha_L B_0$, with $B_0$ computed from the prefill tiers alone, and keeps it enabled while admission is active. To preserve decode headroom, it stops adopting units when requests are waiting for decode memory or running requests occupy most of that memory.

\begin{algorithm}[t]
\small
\caption{\sys's control loop, run every control interval $\Delta$. Parameter values: \S\ref{sec:design-impl}.}
\label{alg:basin}
\begin{algorithmic}[1]
\Require profile $(\kappa,\lambda_c)$; capacities $(C_G,C_H)$; predictor of $\hat W$
\Require interval $\Delta$; margin $m$; thresholds $\boldsymbol\alpha$; lending cap $\phi$; shedding horizon $T_s$; priority rate $r$
\State read $Q_P,Q_D,h,N,\bar c,\bar\lambda$; $W\gets N\bar c$
\State for each new session $i$: predict $\hat W_i$; $p_i\gets a_i+\hat W_i/r$ \Comment{fixed}
\Statex \textit{Capacity}
\State $C\gets\max(C_G,C_H)$; $B_0^0\gets \min(m\kappa x_k C, C)$
\State $\textit{lend}\gets (W\ge \alpha_L B_0^0\vee\textit{admit})\wedge$ spare decode memory
\State $L\gets$ held-unit capacity, at most $\phi$ of decode memory; $C_{\text{eff}}\gets\max(C,C_G+L)$
\State $B_0\gets \min(m\kappa x_k C_{\text{eff}}, C_{\text{eff}})$
\Statex \textit{Entry}
\If{$W\ge B_0$ \textbf{or} $(\textit{admit}\wedge W\ge \alpha_R B_0)$}
  \State $\textit{admit}\gets\textbf{true}$; adapt $B$ to $h$ within a bounded range around $B_0$
\ElsIf{parked queue nonempty}
  \State $\textit{admit}\gets\textbf{false}$; raise $B$ gradually
\Else
  \State $\textit{admit}\gets\textbf{false}$; $B\gets\infty$; update \textit{plateau} from $h$
\EndIf
\Statex \textit{Scheduling and shedding}
\State $\textit{order}\gets$ prefill backlog or decode-memory wait persists
\State $\textit{order}\gets\textit{order}\vee(W\ge \alpha_O B_0)\vee\textit{admit}$
\State order both engine queues by $p$ when \textit{order}
\State admit queue heads by $p$ while they fit $B$ or no residents remain
\If{$\bar\lambda>\lambda_c$ \textbf{and} highest-priority parked wait $>T_s$}
  \State shed the lowest-priority parked session
\EndIf
\State release actions after their conditions clear for a cool-down
\end{algorithmic}
\end{algorithm}

\begin{figure}[t]
  \centering
  \includegraphics[width=\columnwidth]{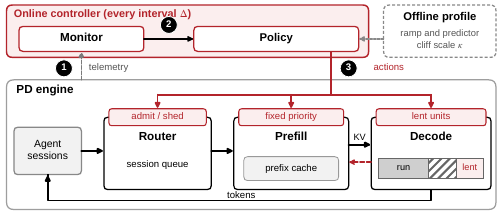}
  \caption{\sys's online controller and offline profile.}
  \label{fig:design}
\end{figure}

\paragraph{Units.} A decode instance retains the prompt state already delivered by the PD transfer as a \emph{unit} after the call completes. The unit contains the transferred KV pages and, for hybrid models, a copy of the prompt-end recurrent state stored in lent pages. On the next turn, prefill compares its local prefix coverage with the unit. If the unit covers more context, decode writes the missing state into prefill's KV pages in layer order, overlapping transfer with execution (\S\ref{sec:design-impl}).

\paragraph{Reclamation.} Decode retains units only with headroom beyond queued and active decode demand: adoption requires spare tokens beyond that demand, and reclamation begins at a lower margin. Lending is capped at a fraction $\phi$ of decode memory. Units also held by prefill are reclaimed first, followed by LRU units. Decode allocations reclaim lent pages before retracting running requests. Reserving decode KV at prefill completion prevents requests awaiting prefill from occupying this headroom.

\subsection{Session Priority for Cache Packing}
\label{sec:design-order}

To fit more sessions within the cache budget, \sys gives smaller predicted footprints a priority advantage. At its first call, session $i$ receives the priority $p_i=a_i+\hat W_i/r$, where $a_i$ is its arrival time and $r$ converts tokens of predicted demand into seconds; smaller values go first. A larger $r$ gives arrival time more weight; a smaller $r$ favors smaller footprints more strongly. \sys uses the same key in the parked queue, the prefill queue and the decode-memory queue. Returning calls retain their key even after eviction, allowing the session to recompute and finish under its original priority. Smaller sessions can precede larger residents; admission controls the active cohort through the entry rate $I(t)$ in Eq.~\ref{eq:entry}.

Keeping the key fixed avoids reshuffling service order as a session waits or its context grows. The arrival term limits overtaking: if a new session $j$ arrives after $a_i+\hat W_i/r$, then $p_j>p_i$ even for a zero-size prediction. Later arrivals therefore stop passing session $i$ after a finite interval; its actual wait also depends on the work already ahead of it. With equal estimates, the key reduces to arrival order.

Ordering engages when a prefill backlog or a wait for decode memory persists. It also engages at $W\ge\alpha_O B_0$ or whenever admission is active (Table~\ref{tab:control}). When the active cohort operates near peak departure capacity, Eq.~\ref{eq:queue} estimates how quickly completions drain the remaining external backlog.

\subsection{Admission for Working-Set Control}
\label{sec:design-residency}

Admission complements the fixed service order by limiting the combined reservations of active sessions, following the working-set principle~\cite{denning1968working}.

Parked and newly arriving sessions form one queue in priority order. With a finite budget $B$, \sys admits its head $i$ when $\sum_{j\in\mathrm{res}}\hat W_j+\hat W_i\le B$. The sum reserves each resident's predicted peak from the start, covering subsequent growth up to that prediction. When the head's reservation exceeds the remaining budget, it stays parked at the gateway, as in ThunderAgent's router port~\cite{thunderagent,llmdthunder}. If no residents remain, the head enters even when its prediction exceeds $B$, ensuring progress. Admitted sessions continue to completion; prediction errors and capacity changes affect subsequent entry through hit-rate feedback.

\paragraph{Activation.} Admission engages when $W\ge B_0$ and remains active until $W<\alpha_R B_0$ has held for the cool-down. The lower release threshold and cool-down keep control active through short-lived drops in demand. The reference $B_0$ determines activation, while the adapted $B$ determines which first calls enter. When control is inactive and no parked sessions remain, the budget is unrestricted.

\paragraph{Budget adaptation and shedding.} Following page-fault-frequency control~\cite{chu1972pff}, \sys raises $B$ when sessions are parked but reuse remains on its healthy plateau, and lowers it when reuse deteriorates. Adaptation stays within a bounded range around $B_0$; downward steps account for measured resident demand, and admitted sessions continue running. Lent capacity remains in the budget during a healthy-hit-rate prefill backlog, allowing the engine to use that cache while computation catches up. After admission is released, gradual budget growth spreads the return of parked sessions over time.

\sys uses shedding to remove persistent excess demand. If the long-run mean arrival rate $\bar\lambda$ exceeds the profiled peak departure rate $\lambda_c$ and the highest-priority parked session has waited $T_s$, \sys sheds the lowest-priority parked session, one per interval (Algorithm~\ref{alg:basin}). Under a size-aware priority the oldest parked session is often a large one that waits by design; the head of the queue waiting $T_s$ shows that admission itself has stalled.

\subsection{Offline Profiling}
\label{sec:design-profile}

\sys locates the cache cliff with a stepped load ramp. At each step, it lets the number of active sessions settle and records working set and hit rate. It fits $\kappa$ by aligning a reference curve, whose knee is $x_k\approx1.9$, with the observed transition from the healthy plateau to the low-reuse floor. Engine rates and session statistics are measured separately, and \sys uses the lower quartile of $\kappa$ across repeated ramps. For several prefill instances, it profiles each router separately to account for prefix distribution, then applies a single budget to aggregate capacity (\S\ref{sec:eval-multip}).

To estimate each session's peak context, \sys fits a separate ridge regression~\cite{hoerl1970ridge} for each workload. The regression predicts log peak context length from term frequency--inverse document frequency (TF-IDF) features~\cite{salton1988termweighting} of the task text and the log lengths of the task text and first prompt. We regularize the text-feature coefficients and fit the two length coefficients freely. Predictions are cross-fitted: each trajectory is evaluated using a model trained on the remaining trajectories, with vocabulary, IDF weights and feature centering computed from that training set. On Terminal-Bench and SWE-bench Pro, these predictions have rank correlations of 0.64 and 0.56 with peak context, and log-error standard deviations of 0.54 and 0.34, respectively.

\subsection{Implementation}
\label{sec:design-impl}

\sys adds a session gateway before SGLang's PD router and extends the engine's allocation and cache paths for unit lending (Fig.~\ref{fig:design}). The hardware prototype implements arrival-time ordering, current-context admission, lending and shedding. The simulator evaluates the complete predictor-driven controller, including peak-context reservations and the fixed size-aware key, on the same calibrated engine model.

\paragraph{Parameters.} The predictor-driven controller uses $\Delta=30$\,s (5\,s while backlog or a control persists), $m=0.9$, $(\alpha_L,\alpha_O,\alpha_R)=(0.8,0.9,0.8)$ for lending, ordering and release, $\phi=0.5$, $T_s=15$\,min and $r=30$\,tokens/s. Section~\ref{sec:eval-ablation} varies these parameters, together with the implementation settings for cool-down, backlog persistence, context smoothing, reopening and lending headroom.

\paragraph{Gateway.} The gateway identifies each call's session, holds parked first calls and runs the controller. It polls existing Prometheus metrics every 5\,s to derive hit rate, outstanding prefill work and requests awaiting decode memory after prefill completion. The hardware prototype uses session arrival times to order both engine queues; the simulator supplies the predicted size-aware key to the corresponding queues. When lending is active, the gateway marks prompts for retention on decode.

\paragraph{Reservation at prefill completion.} Decode KV reservation follows prefill completion, while prefill retains the request's KV until transfer. We extended this allocation path to NIXL, the decode radix cache, write-through HiCache on prefill and hybrid-model state.

\paragraph{Unit lending.} Decode stores completed prompt pages by unit identifier, with Mamba state copied before the first decode step. Allocation and retraction paths adopt and reclaim units under the headroom rules. Prefill reports which units it still holds so decode can reclaim redundant copies first, and maintains a per-session index of the latest unit's decode instance and length.

\begin{figure}[t]
  \centering
  \includegraphics[width=\columnwidth]{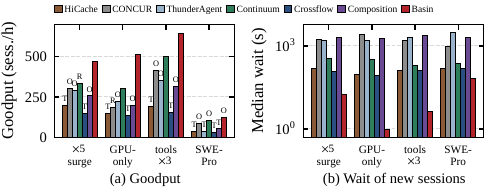}
  \caption{Real H200 1P3D results: three $\times$5 surge scenarios on Terminal-Bench, and the same surge on SWE-bench Pro from 2.6 sessions/min. The prototype uses arrival-time ordering and current-context admission. (a) Goodput; letters above bars give the run state: T trapped, O overloaded, R recovering; unmarked bars are healthy. (b) Median first-call wait.}
  \label{fig:realbars}
\end{figure}

When prefill's local radix match covers fewer tokens than the latest unit, prefill allocates pages and requests NIXL writes of the missing state in layer order. A 25K-token unit transfers within tens of milliseconds over NVLink. To overlap transfer with execution, decode publishes completion flags after each group of layers, and a Triton kernel waits on the corresponding flag before each layer executes. Attention and Mamba layers run outside SGLang's captured piecewise-prefill segments, keeping the captured graphs intact. The wait kernel is launched only for batches fetching unit data. A CPU monitor detects timeouts or failed decode instances and falls back to recomputation.

\setcounter{topnumber}{\basinSavedTopNumber}

\edef\basinEvalSavedTopNumber{\number\value{topnumber}}
\setcounter{topnumber}{1}

\section{Evaluation}
\label{sec:evaluation}

\subsection{Methodology}
\label{sec:eval-method}

Hardware experiments compare \sys with existing policies on H200 and H100 (\S\ref{sec:eval-real}). Simulation covers the broader deployment, load and policy sweeps. The comparisons in \S\ref{sec:eval-real} report real measurements; the subsequent performance sweeps use simulation. Section~\ref{sec:eval-sim} reports calibration and hardware agreement.

\paragraph{Simulator and deployments.} The closed-loop simulator uses separate H100 and H200 calibrations (\S\ref{sec:eval-sim}). It accounts for token-position-dependent prefill cost, batch- and context-dependent decode steps, KV reservation and retraction, recurrent-state slot limits, and host and decode-lent cache tiers.

We simulate 1P1D, 1P2D and 1P3D, plus 2P6D in \S\ref{sec:eval-multip}. Unless specified as GPU-only, HiCache has a host tier twice the GPU prefix capacity. Base load is relative to the deployment's profiled critical rate. The main comparison starts at half the deployment's profiled critical rate and applies a $\times$5 surge during minutes 15--35 on 1P1D and 1P3D, using the default workload, tool time scaled by 3, or GPU-only caching. Further sweeps vary trigger size and duration, output length, cache capacity and routing.

\paragraph{Workloads.} We replay the Terminal-Bench trajectories of \S\ref{sec:problem} and 70 SWE-bench Pro~\cite{swebenchpro} sessions as a second coding-agent workload on hardware. On Terminal-Bench and SWE-bench Pro, median prompts contain 25.6K and 83K tokens, median peak contexts reach 40K and 107K tokens, and median tool gaps are 0.42\,s and 48\,ms, respectively. Each call becomes eligible after its predecessor completes and the recorded tool gap elapses. The deployment and policy sweeps use Terminal-Bench unless specified; sensitivity experiments scale tool time or output length, and Bailian supplies alternative arrival patterns for these trajectories.

\paragraph{Real deployments.} We implement the arrival-time \sys prototype, Continuum, ThunderAgent and CONCUR in SGLang with NIXL PD transfers and page size 64 (\S\ref{sec:design}). The hardware deployments are one four-H200 node in 1P3D and two H100 NVL GPUs in 1P1D. Both serve Qwen3.6-35B-A3B in FP8, with 30 linear-attention layers among its 40 layers. The replay client uses the same session trajectories, arrival schedules and tool gaps as the corresponding simulated configuration. Runs last 2\,h, or 2.5\,h with tool time scaled by 3; session metrics cover arrivals in the first 90 minutes.

\paragraph{Baselines.} We compare HiCache alone with concurrency control in CONCUR~\cite{concur} and ThunderAgent~\cite{thunderagent,llmdthunder}, ordering and KV retention in Continuum~\cite{continuum}, and decode-side elastic prefill in Crossflow~\cite{crossflow}. A \emph{composition} combines ThunderAgent's admission bound, Crossflow's borrowing, and Continuum's ordering and TTL pinning. CONCUR is included in the main end-to-end and hardware comparisons and in the steady-load and production-arrival sweeps; each later figure shows the systems its comparison concerns, in the same order and colors.

ThunderAgent bounds admitted footprint by the larger cache tier. Our CONCUR port uses decode occupancy for KV usage and enables turn-boundary pausing. We raise the hit-rate threshold from 0.2 to 0.7 to place it within the observed range. Scheduling priorities apply to both prefill and decode-memory queues. Crossflow shares the mixed-step token budget between prefill and decode and adapts its lease to a 50\,ms ITL SLO. \sys uses the same SLO and is configured as in \S\ref{sec:design}.

\begin{figure}[t]
  \centering
  \includegraphics[width=\columnwidth]{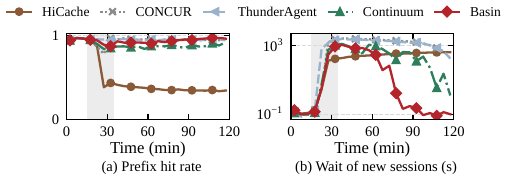}
  \caption{The $\times$5 surge (minutes 15--35, shaded) on the real H200 1P3D deployment, in 5-minute bins of arrival
  time: (a) prefix hit rate of the calls; (b) median wait of the sessions arriving in the bin.}
  \label{fig:real}
\end{figure}

\paragraph{Metrics.} Goodput is the number of completed sessions per hour with slowdown at most 4 and at least 95\% of decode steps meeting the ITL SLO. Slowdown divides elapsed session time, including first-call wait, by an unloaded reference computed from tool gaps, call overheads, new-suffix prefill and batch-one decode using the engine's calibrated costs. Rejected and shed sessions count as failed, and completion fraction covers all evaluated arrivals. Warm-start sessions are admitted initially and excluded from session metrics. Latency metrics include median and p90 TTFT, first-call wait and session slowdown.

A 20-minute window is \emph{degraded} if median TTFT exceeds 5\,s, median first-call wait exceeds 5 minutes, or at least 25\% of arriving sessions never start. A run is \emph{trapped} if its final two windows remain degraded, improve by less than 25\%, and both show prefix hit rate at least 0.10 below the pre-trigger rate. These windows end at least 35 minutes after the trigger. Persistent degradation without improvement but with retained cache reuse is labeled \emph{overloaded}; an improving run is \emph{recovering}. Labels are unchanged for hit-rate thresholds from 0.10 to 0.20.

\begin{figure}[t]
  \centering
  \includegraphics[width=\columnwidth]{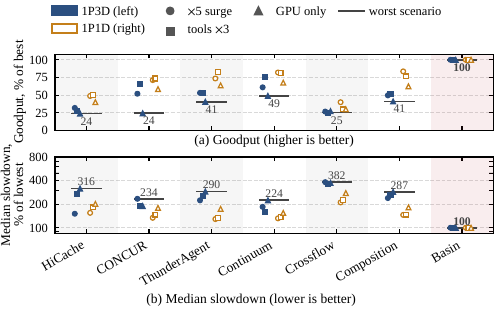}
  \caption{\simtag Each marker is one scenario on 1P3D or 1P1D: a $\times$5 surge, the same surge with tool time
  scaled by 3 or a GPU-only cache. Color marks the deployment (1P3D filled, left; 1P1D hollow, right), shape the scenario, and bars each system's worst scenario.
  (a) Goodput as a percentage of the best system in the scenario. (b) Median slowdown of completed sessions,
  including the wait before the first call, as a percentage of the lowest in the scenario.}
  \label{fig:e2e}
\end{figure}

\subsection{Real Deployments}
\label{sec:eval-real}

The hardware prototype uses arrival-time ordering and current-context admission (\S\ref{sec:design-impl}); the predictor-driven controller is evaluated in simulation from \S\ref{sec:eval-e2e} on. On hardware, \sys achieves 1.20--1.70$\times$ the goodput of Continuum and 2.35--3.59$\times$ that of HiCache alone across the four H200 1P3D scenarios (Fig.~\ref{fig:realbars}). HiCache remains trapped in all four, completing only 0.28--0.54 of sessions. On Terminal-Bench its prefix hit rate falls from 0.97 to 0.33--0.35. Crossflow is trapped in all four as well, completing 0.26--0.53 of sessions. On Terminal-Bench, ThunderAgent and the composition maintain reuse but impose median first-call waits of 26--39 minutes; on SWE-bench Pro both are trapped. CONCUR also preserves reuse except with GPU-only caching, where its hit rate falls to 0.33 before its concurrency window contracts and recovers slowly. On SWE-bench Pro, whose sessions reach a median of 107K tokens against 40K on Terminal-Bench, \sys completes 126.7 sessions/h against 106.0 for Continuum and 86.0 for CONCUR, with a median first-call wait of 68\,s.

\begin{figure}[!t]
  \centering
  \includegraphics[width=\columnwidth]{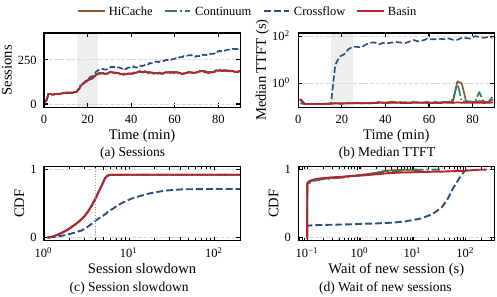}
  \caption{\simtag Response to a $\times$3 surge during minutes 15--25 (shaded) on 1P3D with output lengths
  scaled by 3, where decode saturates first: (a, b) time series and (c, d) session distributions.
  Unfinished sessions appear at the right edge of (c).}
  \label{fig:dynamics}
\end{figure}

Across the three Terminal-Bench scenarios, Continuum avoids the trap, but its hit rate remains 10--15 points below the pre-surge rate and new sessions wait 3--6 minutes. Under \sys, median first-call wait is at most 18\,s. The GPU-only scenario gives the largest gap to Continuum, because lending provides the only cache tier beyond prefill GPU memory.

In the default surge (Fig.~\ref{fig:real}), HiCache's hit rate collapses within ten minutes of surge onset and remains low through the end of the run. \sys drains the backlog by minute 75, while Continuum drains it near the end of the run.

On H100 1P1D, decode saturates first (Eq.~\ref{eq:condition}). Under the $\times$5 surge, \sys, Continuum and the composition achieve 56.7, 55.3 and 55.3 sessions/h. For the arrival-time prototype, limited lending headroom and the absence of cache-driven vulnerability leave little throughput benefit over ordering alone; the size-aware priority adds 1.20--1.48$\times$ in simulation (\S\ref{sec:eval-e2e}).

\begin{figure}[t]
  \centering
  \includegraphics[width=\columnwidth]{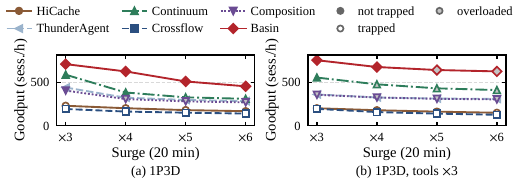}
  \caption{\simtag Surges of increasing size, 20 minutes long. End-state markers follow the legend and \S\ref{sec:eval-method}.}
  \label{fig:sweep}
\end{figure}

\begin{figure}[!t]
  \centering
  \includegraphics[width=\columnwidth]{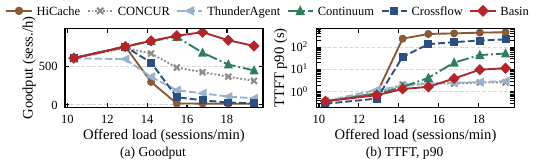}
  \caption{\simtag Steady load on 1P3D with tool time scaled by 3.}
  \label{fig:load}
\end{figure}

\subsection{Surge Response}
\label{sec:eval-e2e}

\sys achieves the highest goodput in all six scenarios of Fig.~\ref{fig:e2e}. Each baseline loses at least 51\% in one or more scenarios. On cache-first 1P3D, \sys exceeds Continuum, the strongest baseline in all three cases, by 1.64$\times$ under the default surge, 1.33$\times$ with longer tools and 2.05$\times$ with GPU-only caching. Relative to HiCache, the gain is 3.13--4.17$\times$. Median slowdown falls from 6.6 to 3.6 under the default surge and from 6.6 to 3.0 with GPU-only caching.

On 1P1D, \sys achieves 1.20--1.48$\times$ the strongest baseline's goodput. Under the default surge, it completes 92\% of sessions and reduces p90 first-call wait to 0.56 of the composition's.

\begin{figure}[t]
  \centering
  \includegraphics[width=\columnwidth]{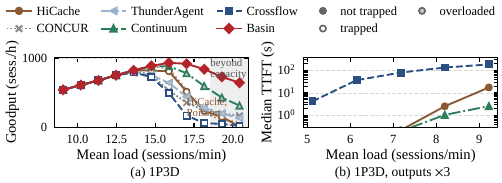}
  \caption{\simtag Session arrivals shaped by the Bailian coder trace, two hours.
  Shaded: offered load beyond capacity. (a) Dotted: HiCache under Poisson arrivals; marker fill as in Fig.~\ref{fig:sweep}. (b) Median TTFT in the second hour, where the first call of a parked session includes its wait at the router.}
  \label{fig:realarr}
\end{figure}

Fig.~\ref{fig:dynamics} follows a surge on 1P3D with output length scaled by 3, making decode the first bottleneck. \sys, HiCache and Continuum achieve 232, 231 and 229 sessions/h, respectively, with p90 first-call wait of 4\,s under \sys. Crossflow's median TTFT rises to about 100\,s as prefill competes with decoding.

In 15 healthy control scenarios, \sys retains at least 99.6\% of the best baseline's goodput. At most 1.3\% of its sessions violate the ITL SLO on more than 5\% of their steps across the main end-to-end scenarios.

\subsection{Trigger Size and Sustained Load}
\label{sec:eval-robust}

Fig.~\ref{fig:sweep} varies 20-minute surge factors from 3 to 6 on 1P3D and 1P3D with tool time scaled by 3. Their base loads are 0.55 and 0.66 of the profiled critical rate, respectively. Crossflow remains trapped throughout this range. Under \sys, runs are healthy or recovering at the end, with 2.71--4.19$\times$ HiCache's goodput and 1.21--1.63$\times$ the strongest baseline's. At least 95\% of sessions complete, compared with 32--48\% under HiCache, while at most 2.2\% are shed. Admission-bound systems retain much longer queues: sessions arriving 35 minutes after the two largest surges still wait about 35--48 minutes under ThunderAgent and the composition.

In the steady-load sweep (Fig.~\ref{fig:load}), \sys peaks at 948 sessions/h at an offered rate of 16.8 sessions/min and retains 769 at 19.4 sessions/min, the highest load shown. Continuum, CONCUR and ThunderAgent achieve 449, 312 and 84 sessions/h there. HiCache completes fewer than 20 sessions/h from 15.5 sessions/min onward. At the highest load, p90 TTFT is 11\,s under \sys, compared with 53\,s for Continuum and 225\,s for Crossflow. ThunderAgent keeps p90 TTFT at 3.1\,s by holding sessions at admission, trading throughput for shorter admitted-call queues.

Replaying Bailian's arrival patterns with Terminal-Bench session trajectories produces the same goodput advantage. At 18.2 sessions/min under the coder pattern, just below the profiled critical rate of 18.5, \sys achieves 1.41$\times$ Continuum's goodput, 2.98$\times$ ThunderAgent's and 3.4$\times$ CONCUR's. At the highest mean load, 20.4 sessions/min, \sys achieves 645 sessions/h under the coder pattern and 724 under trace A, compared with 315 and 428 for Continuum and 169 for CONCUR under the coder pattern (Fig.~\ref{fig:realarr}).

\begin{figure}[!t]
  \centering
  \includegraphics[width=\columnwidth]{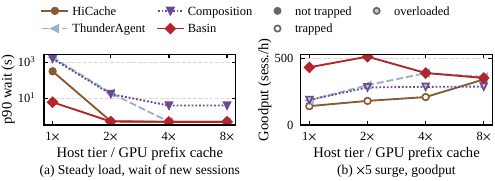}
  \caption{\simtag ThunderAgent, Composition and \sys under increasing host-cache capacity on 1P3D. Marker fill as in Fig.~\ref{fig:sweep}.}
  \label{fig:hostr}
\end{figure}

With output lengths scaled by 3, the profiled critical rate is 9.9 sessions/min. At the highest coder load, 9.2 sessions/min, \sys's second-hour median TTFT is 0.2\,s, compared with 17.8\,s for HiCache and 187.8\,s for Crossflow. Across the two highest loads, its p90 first-call wait is 20--34 minutes, compared with 45--54 under ThunderAgent and 46--59 under CONCUR. Maintaining reuse shifts part of the delay to admission; session slowdown includes both components.

\begin{figure}[t]
  \centering
  \includegraphics[width=\columnwidth]{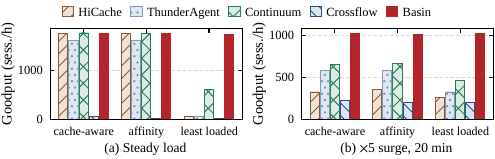}
  \caption{\simtag 2P6D under three prefill routers.}
  \label{fig:multip}
\end{figure}

\subsection{Sources of Performance Differences}
\label{sec:eval-analysis}

\emph{Resource borrowing.} Crossflow runs prefill on decode instances, sharing compute with token generation. \sys lends cache memory while keeping prefill computation on prefill instances, with decode allocations taking priority over lent pages (\S\ref{sec:design-capacity}). Across the 1P1D scenarios, Crossflow achieves 0.30--0.40 of the best system's goodput (Fig.~\ref{fig:e2e}).

Crossflow is also sensitive to per-chunk overhead. Charging a fixed 20\,ms per prefill chunk reduces its goodput from 206 to 174 sessions/h under a $\times$3 surge and from 76 to 5 in the high-load tool-time-scaled case, relative to zero fixed charge. \sys's goodput is unchanged in this sensitivity test.

\emph{Admission signals.} In our CONCUR port, decode KV usage covers requests reserved, transferring or running on decode. Tool-idle sessions retain prefix-reuse demand without an in-flight decode reservation. At equal decode occupancy, different numbers of tool-idle sessions impose different cache demand, and the hit-rate trigger responds only after reuse deteriorates. CONCUR achieves 0.52 of \sys's goodput under the default 1P3D surge and 0.24 with GPU-only caching.

\emph{Admission threshold.} ThunderAgent ties its bound to the larger prefill cache tier. With a host tier equal to GPU capacity, \sys achieves 2.33$\times$ ThunderAgent's goodput and 2.30$\times$ the composition's (Fig.~\ref{fig:hostr}). At host tiers of four and eight times GPU capacity, the ratios narrow to 0.98--1.00 against ThunderAgent and 1.22--1.36 against the composition. At this cache size, decode limits service: at eight times GPU capacity \sys completes 14\% more sessions within the slowdown SLO than the earlier arrival-time controller and 20\% more than ThunderAgent, but 35\% of them exceed the ITL allowance, against 18\% and 24\%. The cache-based budget allows decode concurrency to grow beyond the ITL target.

\begin{figure}[!t]
  \centering
  \includegraphics[width=\columnwidth]{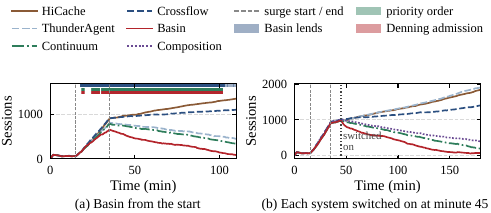}
  \caption{\simtag A $\times$5 surge for 20 minutes (minutes 15--35) on 1P3D; sessions include those parked at the router. (a) All systems; the
  bands at the top mark when \sys lends decode memory, orders queues by its priority and limits admission. (b) Each
  system switched on at minute 45, after HiCache alone is trapped.}
  \label{fig:recovery}
\end{figure}

\emph{Eviction policy.} Across surge factors 3--6, next-call-oracle eviction achieves 1.08--1.18 of LRU's goodput and 0.28--0.41 of \sys's. Tool-gap pinning remains within 1\% of LRU. Largest-first eviction yields substantially lower goodput even in the steady controls.

\begin{figure}[!t]
  \centering
  \includegraphics[width=\columnwidth]{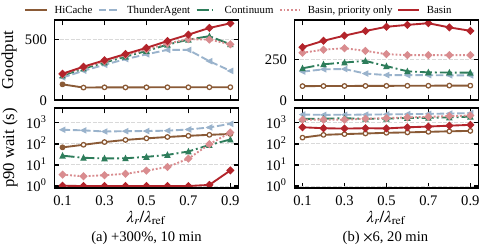}
  \caption{\simtag GPU-only 1P3D under varying post-surge arrivals: goodput (top, sessions/h) and p90 first-call wait (bottom, s; below 1\,s drawn at 1\,s) after (a) a $+300\%$ surge for 10 minutes and (b) a $\times6$ surge for 20 minutes. Arrivals before the surge are $0.5\,\lambda_{\mathrm{ref}}$; results average three seeds. Hollow markers: $L_{\mathrm{late}}>10$\,s in every seed.}
  \label{fig:recsys}
\end{figure}

\subsection{Multiple Prefill Instances}
\label{sec:eval-multip}

Routing changes usable cache capacity (Fig.~\ref{fig:multip}). On 2P6D, the profiled critical rates are 39 sessions/min with cache-aware routing, 41 with affinity and 29 with least-loaded routing. The router-specific profiles give lower-quartile cliff scales of 0.81, 1.38 and 0.55, respectively. Under a $\times$5 surge, \sys exceeds the strongest baseline by 1.58$\times$, 1.53$\times$ and 2.18$\times$, shedding at most 1.2\% of sessions. Least-loaded routing scatters prefixes across instances: in its steady control, \sys achieves 1742 sessions/h, while HiCache and ThunderAgent fall to 53 and 58.

\begin{figure}[t]
  \centering
  \includegraphics[width=\columnwidth]{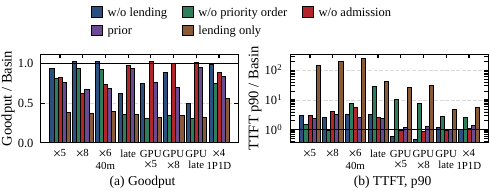}
  \caption{\simtag Ablation of \sys's actions, relative to the full \sys, on 1P3D with a host tier
  and with a GPU-only cache (``GPU''). ``40m'': a 40-minute surge; ``late'': \sys switched on at minute 45 in a
  trapped engine; ``1P1D $\times$4'': a $\times$4 surge on 1P1D; ``prior'': the earlier controller, with arrival-time ordering and current-context admission.}
  \label{fig:ablation}
\end{figure}

\subsection{Recovering a Trapped Engine}
\label{sec:eval-recovery}

Fig.~\ref{fig:recovery}b activates each policy at minute 45, after HiCache is trapped. \sys reduces the total number of resident and parked sessions to at most 60 by 103--128 minutes after activation across five seeds, completing 99.9\% of measured sessions. Its goodput is 301.1 sessions/h, compared with 167.8 for Continuum, 136.0 for the composition and 108.9 for HiCache. HiCache completes 36.1\% of sessions. The long drain time reflects the accumulated backlog even after service improves; Fig.~\ref{fig:recovery}a shows the corresponding run with controls active before the surge.

\subsection{Systems under Post-Surge Arrivals}
\label{sec:eval-boundary}

Fig.~\ref{fig:recsys} applies the same post-surge arrival intervention to the evaluated systems on GPU-only 1P3D and 1P3D with a host tier. It uses the evaluation's engine options while retaining the reference rates of \S\ref{sec:model-boundary}. Goodput and first-call wait use the metrics of \S\ref{sec:eval-method}; the late-latency marker uses $L_{\mathrm{late}}$ from \S\ref{sec:characterization}.

The systems differ in both departure capacity and the delay imposed on new sessions. Following the $+300\%$ surge, ThunderAgent's p90 first-call wait is 275--467\,s at post-surge rates up to half the reference rate, while \sys stays below 17\,s across the sweep. Following the larger surge, Continuum leaves new sessions waiting 1.4--2.0K\,s. Across GPU-only and host-tier settings, \sys achieves 1.00--1.38$\times$ Continuum's goodput after the $+300\%$ surge and 1.29--2.77$\times$ after the $\times6$ surge, and 1.09--3.08$\times$ ThunderAgent's. The larger-surge gain over Continuum is 1.66--2.77$\times$ with GPU-only caching and 1.29--2.10$\times$ with the host tier. Without lending and admission, the priority alone retains 0.95--1.05 of Continuum's goodput after the $+300\%$ surge and reaches 1.19--1.64$\times$ after the $\times6$ surge: the size-aware priority and lent capacity both contribute to the gain over program ordering.

\begin{figure}[t]
  \centering
  \includegraphics[width=\columnwidth]{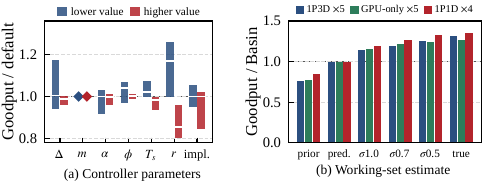}
  \caption{\simtag Sensitivity of Basin's goodput (five seeds). (a) One controller parameter set lower or higher, relative to the default: $\Delta$ 15/60\,s, $m$ 0.7/1.0, $\boldsymbol\alpha$ 0.5--0.7/0.95--1 (three thresholds pooled), $\phi$ 0.3/0.7, $T_s$ 7.5/30\,min, $r$ 10/100\,tokens/s, and the nine implementation settings (``impl.''). Bars span the five ablation triggers and an hour of sustained overload on 1P3D; the white tick marks the median. $m$ leaves the bound unchanged (diamond), as the capacity term binds. (b) Relative to the predictor-driven controller: the earlier arrival-time controller (``prior''), and the new controller supplied with true peaks perturbed by lognormal error $\sigma$ or with exact peaks.}
  \label{fig:sensitivity}
\end{figure}

\subsection{Mechanism Ablations}
\label{sec:eval-ablation}

Fig.~\ref{fig:ablation} removes each mechanism and includes lending alone. Returning to the earlier arrival-time controller retains 0.67--0.77 of the full policy's goodput under large surges on 1P3D, 0.84 on 1P1D and 0.94--0.95 during recovery. This comparison changes both service order and demand accounting. To isolate engine ordering, we keep predicted admission but remove the priority from both engine queues: goodput falls to 0.81--0.95 with the host tier, 0.31--0.34 with GPU-only caching, 0.36 during recovery and 0.55 under sustained overload. Removing lending retains 0.94--1.03 with the host tier and 0.50--0.89 in the GPU-only and recovery cases. Lending alone achieves 0.32--0.56.

Admission matters where the cache tier is large: without it, \sys retains 0.62--0.82 under host-tier surges and 0.90 on 1P1D, while the GPU-only, recovery and sustained-overload runs stay within 1.00--1.03. Bounding at the profiled cliff alone, without the capacity term, retains 0.82--0.92 with the host tier, and dropping lent capacity from the bound under a compute backlog retains 0.83--0.87 with GPU-only caching. Under sustained overload with the host tier, admission reduces the mean number of active sessions from 852 to 252 and the prefill queue from 620 to 14 during minutes 60--75; it sheds 82 sessions and improves goodput by 1.25$\times$, or 1.08$\times$ with GPU-only caching.

An always-on variant achieves 1.05--1.11$\times$ the adaptive controller's goodput under the two largest finite surges but 0.70--0.92 under steady and sustained load, where it holds sessions the adaptive controller admits. Scaling the profiled cliff by 0.5--2$\times$ changes goodput by at most 1\%, as the capacity term sets the bound.

\paragraph{Controller parameters.} We vary the controller parameters of \S\ref{sec:design-impl} to a lower and a higher value, one at a time, on the five ablation triggers and an hour of sustained overload on 1P3D (five seeds). Of the 204 variant--scenario pairs, 81\% stay within 3\% of the default goodput (median change 0.5\%), and all lie within 0.80--1.33. The margin $m$ has no effect in these sensitivity configurations, where the capacity term sets $B_0$. Each component of $\boldsymbol\alpha$ changes goodput by at most 8\%. A shorter interval ($\Delta=15$\,s) raises goodput by up to 1.17$\times$ under sustained overload and costs up to 6\% under the $\times$8 surge; 60\,s costs up to 4\%. The lending cap $\phi$ changes goodput by 0.97--1.07$\times$, and $\phi=0.7$ lowers the GPU-only p90 first-call wait to 0.77$\times$. A shorter shedding horizon ($T_s=7.5$\,min) raises goodput by up to 1.07$\times$ under sustained overload by shedding earlier; a longer one (30\,min) lowers it to 0.93$\times$. The priority rate $r$ trades goodput against tail admission delay. At $r=10$\,tokens/s, a stronger preference for small footprints raises goodput by 7--33\%, while p99 first-call wait rises by 19--43\%. At $r=100$, closer to arrival order, goodput falls to 0.80--0.96$\times$. We use $r=30$ to balance goodput and tail wait. Among the implementation settings, a wide budget range ($[0.25,4]\,B_0$) costs up to 16\% under sustained overload, slower context smoothing (90\,min) up to 12\% under the $\times$8 surge, and doubled budget steps up to 7\% on 1P1D; the others stay within 0.96--1.03 across their tested ranges (Fig.~\ref{fig:sensitivity}a).

\paragraph{Working-set estimate.} Fig.~\ref{fig:sensitivity}b varies predictions within the new controller and includes the earlier arrival-time controller as a reference. The latter reaches 0.76--0.84 of the full policy's goodput in the three plotted scenarios and 0.67--0.95 across ablation triggers. Supplying exact peak contexts raises goodput by 1.27--1.35$\times$ in the plotted scenarios and 1.10--1.51$\times$ across triggers, showing further benefit from improved predictions. Exact peaks perturbed by independent lognormal error with $\sigma=1.0$ also outperform the predictor (1.15--1.19$\times$ in the plotted scenarios), indicating that error structure and session ranking matter alongside error magnitude. Under sustained overload, the variants reach 0.98--1.00 of the default goodput.

The first-call features of \S\ref{sec:design-profile} give leave-one-trajectory-out rank correlations of 0.64 on Terminal-Bench and 0.56 on SWE-bench Pro. Additional prediction checks on GitHub Copilot production traces~\cite{liu2026agentic} and two Bailian traces train on earlier sessions and test on later ones: first-call prompt-segment lengths and model label give a rank correlation of 0.60 on Copilot, and first-request length gives 0.71--0.74 on Bailian. Log-error standard deviations across these checks range from 0.34 to 0.69.

\subsection{Simulator Fidelity}
\label{sec:eval-sim}

These comparisons use the earlier arrival-time controller for \sys, matching the hardware implementation; the engine and cache models are shared with the predictor-driven sweeps.

\paragraph{H100.} Calibration covers five 1P1D configurations: surge runs of HiCache, Continuum, ThunderAgent and Crossflow, plus the GPU-only reproduction. Across five simulated seeds per configuration, mean goodput differs from the corresponding real run by at most 8.9\%. All five real runs and their seed-0 simulations are classified healthy. In the later three-system H100 comparison, goodput error is within 3\%. The simulator underestimates \sys's ITL-violating session share, at 0.5\% against 13.6\% measured.

\paragraph{H200.} We fit decode-step cost to batch size and running context using two-minute windows from eighteen calibration runs, including the main trace and SWE-bench Pro. The latency fit has RMSE 0.57\,ms. Prefill rate and the SWE-Pro prompt-length scale are also calibrated. Under the evaluation metrics, simulated goodput averaged over five seeds differs from the fifteen main-trace 1P3D runs in this calibration set by 8.8\% on average and at most 32\%. The largest \sys error is a 21\% underestimate in the GPU-only case. For the earlier arrival-time controller used by the hardware prototype, simulated \sys/Continuum goodput ratios of 1.18--1.61 are below the measured 1.29--1.70.

We assess accuracy beyond calibration using thirteen additional H200 1P3D runs: Crossflow and the composition in the three Terminal-Bench scenarios, and all seven systems on SWE-bench Pro. These runs were excluded from fitting. For the six Terminal-Bench runs, simulated goodput is within 2--12\% of the measurement. On SWE-bench Pro it is within 5--11\% for HiCache, CONCUR, Continuum and \sys, which it ranks as measured, but it overestimates ThunderAgent and Crossflow by 33\% and 37\%, respectively. For the composition, it predicts 2.6$\times$ the measured goodput.

\paragraph{Transfer-cost sensitivity.} Replacing the fixed 10\,ms PD transfer charge with a prompt-proportional charge changes \sys's goodput by at most 1.4\% across five comparisons.

\setcounter{topnumber}{\basinEvalSavedTopNumber}

\section{Related Work}
\label{sec:related}

\paragraph{Metastability and overload control.} Metastable failures~\cite{bronson2021metastable,huang2022metastable,habibi2024msf,isaacs2025analyzing,dong2022metastability} belong to a broader class of feedback-driven performance failures, including congestion collapse~\cite{jacobson1988congestion}, receive livelock~\cite{mogul1997livelock}, bistable loss networks~\cite{carleial1975bistable,gibbens1990bistability}, thrashing~\cite{denning1968thrashing} and cache stampedes~\cite{nishtala2013memcache,vattani2015stampede}. Overload controllers limit work in progress through admission, dropping or degradation~\cite{welsh2001seda,elnikety2004admission,dagor,breakwater,topfull,protego,atropos,defcon}. Related KV-cache planning uses working sets for capacity sizing~\cite{kvset}. \sys applies the working-set principle~\cite{denning1968working} and page-fault-frequency control~\cite{chu1972pff} to control the number of active sessions.

EfficientAgent relates queueing to reuse distance in a fixed agent pool and sizes the host tier and writes accordingly~\cite{efficientagent}. With open session arrivals, slow service keeps more sessions active and enlarges their aggregate context, allowing cache thrashing to outlive its trigger. We describe this coupling with a population-dynamics approximation~\cite{kurtz1970ode,dai1995fluid} for a partly-open workload~\cite{schroeder2006open}.

\paragraph{Capacity management.} Batching and memory management improve per-GPU throughput~\cite{orca,vllm,sglang,sarathi}, PD disaggregation sizes prefill and decode independently~\cite{distserve,splitwise,mooncake,dynamo}, and cluster managers add, migrate or switch instances with demand~\cite{llumnix,serverlessllm,blitzscale,dynamollm,arrow}. Elastic designs lend idle decode resources to prefill~\cite{crossflow,kairos}. KV systems expand capacity across host, storage and disaggregated memory~\cite{lmcache,sglangHiCache,mooncake,symphony}, retain conversation state~\cite{cachedattention,pensieve}, reduce loading and reuse costs~\cite{cachegen,cacheblend}, and adapt routing or eviction to prefix locality~\cite{preble,marconi,kvcachewild}. Recent work also studies the effectiveness of replacement policies for LLM prefix reuse~\cite{fancyeviction}, and Yao et al.\ derive the LRU hit ratio of multi-turn conversations~\cite{yaolru}. Our model explains how added cache changes vulnerability and identifies when decode becomes the first limit.

AMPD and Dynamo retrieve session KV from the decode worker that produced it~\cite{ampd,dynamoagentic}, while MemServe pools memory across instances for context caching~\cite{memserve}. ConServe places a conversation's later turns on the decode instance holding its state~\cite{conserve}. \sys retains prompts in idle decode memory under explicit headroom and reclamation rules, with the cache index on prefill and running decode requests taking allocation priority.

\paragraph{Agent scheduling and admission.} Prior systems expose program structure~\cite{parrot,teola,pie,infercept}, schedule programs~\cite{autellix,fastserve,vtc,qlm}, and retain or prioritize agent KV~\cite{continuum,coda,kvflow}. Continuum's age order is the case of \sys's fixed priority in which all working-set estimates are equal. Length-aware request schedulers use output-length predictions for batching and memory allocation~\cite{zheng2023response,jin2023s3}, or learn relative lengths to approximate shortest-job-first scheduling~\cite{fu2024ltr}. \sys predicts each session's peak context across turns and uses this estimate to assign a fixed session priority and reserve cache capacity at admission.

TOPAS coordinates prefix residency with request admission~\cite{topas}. Admission policies use predicted latency~\cite{mooncake}, KV usage and hit rate~\cite{concur}, or GPU and tier capacity~\cite{thunderagent,llmdthunder}. Ao et al.\ derive request-level admission from a fluid model with a unique equilibrium~\cite{ao2025fluid}. Our working-set model explains how these controls affect session accumulation and backlog drainage under continuing arrivals.

\paragraph{Instabilities in agent serving.} Prior work studies middle-phase cache thrashing in offline agent batches~\cite{concur}, collapse under sustained agent load~\cite{metronome}, instability from KV growth~\cite{ao2026eviction}, and feedback between hit-rate collapse and re-prefill~\cite{vamp}. We isolate how continuing session arrivals sustain population--cache feedback after a finite trigger, identify resource conditions for this vulnerability, and show that changing subsequent arrivals alone reverses recovery. The post-surge intervention holds the triggering state fixed, establishing the role of continuing load in sustaining failure.

\section{Discussion and Limitations}
\label{sec:discussion}

\paragraph{Generality.} The feedback follows from closed-loop sessions, continuing arrivals and context-dependent miss cost. The population balance and resource comparison apply to full-attention and hybrid models through architecture-specific hit-rate curves and engine bounds; the common per-token KV size cancels from the memory comparison (Eq.~\ref{eq:condition}). In colocated deployments, prefill and decode share compute and memory; the corresponding model bounds would reflect that shared resource pool.

\paragraph{Capacity changes.} \sys assumes fixed capacity between profiles. Adding~\cite{tokenscale}, switching or losing an instance changes the cliff, with new cache capacity becoming useful only as it warms. Hit-rate feedback adapts the budget between profiles, and reprofiling updates the cliff estimate.

\paragraph{Dependent sessions.} The evaluated trajectories consist of independent sessions. Parent--child workflows couple their lifetimes: a parent can retain context while its child waits for service. When sessions share cached pages, those pages contribute once to their aggregate footprint. These workflows therefore call for dependency-aware ordering, admission and prefix accounting.

\paragraph{Decode-first deployments.} The arrival-time hardware prototype achieves goodput similar to Continuum and the composition on the real H100 1P1D surge (\S\ref{sec:eval-real}): with decode setting the departure rate, protecting the cache adds little throughput, and ordering by arrival only redistributes waiting time between resident and new sessions. The size-aware priority still helps there. In simulation it reaches 1.20--1.48$\times$ the strongest baseline's goodput on 1P1D (\S\ref{sec:eval-e2e}), because packing smaller sessions first completes more of them within the slowdown SLO under the same decode and cache capacity.

\paragraph{Evaluation scope.} All hardware experiments serve the hybrid MoE model Qwen3.6-35B-A3B on SGLang. Hardware interventions identify the sustaining feedback, and H100 and H200 runs compare selected operating points. Vulnerability maps, the 36-deployment model comparison, post-surge arrival interventions, 2P6D and broader policy sweeps use simulation. SWE-bench Pro supplies a second hardware workload; tool-time and output-length scaling vary Terminal-Bench trajectories, while Bailian varies their arrival process (\S\ref{sec:eval-sim}).

\section{Conclusion}

Population--cache feedback gives PD-disaggregated agent serving a load-dependent recovery boundary. A finite surge can cross this boundary and leave service degraded at a base rate the engine previously sustained. Real-system interventions show that accumulated active sessions sustain the failure, and controlled simulations show that changing only subsequent arrivals changes recovery. The fluid model identifies resource conditions for vulnerability and explains how entry control and departure capacity change the feedback. \sys applies these dynamics through a fixed session priority, admission by predicted peak context, decode-memory lending and shedding. On real H200 1P3D deployments, the arrival-time prototype achieves 1.20--1.70$\times$ Continuum's goodput and 2.35--3.59$\times$ HiCache's.

{\footnotesize \bibliographystyle{acm}
\bibliography{refs}}
\end{document}